\documentclass[11pt,a4paper]{article}

\usepackage[T1]{fontenc}
\usepackage[utf8]{inputenc}
\usepackage{amsmath,amsfonts,mathtools,amssymb}
\usepackage{graphicx}
\usepackage{tikz}
\usepackage{pgfplots}
\usepackage{pgfplotstable}
\pgfplotstableset{fixed zerofill,precision=3}
\pgfplotsset{compat=1.11}
\pgfplotsset{prefix=plot/}
\usetikzlibrary{decorations,arrows,positioning,patterns,plotmarks,shapes}

\usepackage{multibib}
\newcites{note}{References added to this version}
\usepackage[lining]{ebgaramond} 

\DeclareMathAlphabet\mathboldsf{OT1}{cmss}{bx}{n}
\DeclareMathAlphabet\mathitsf{OT1}{cmss}{m}{it}

\newcommand\dd{\,\mathrm{d}}

\DeclareMathOperator\dirac{\delta}
\DeclareMathOperator\ddirac{\delta'}

\newcommand\e{\mathrm{e}}
\newcommand\gscat[1][1]{g\sb{#1}}
\newcommand\TLgscat[1][1]{\TL{g}\sb{#1}}
\newcommand\TFLgscat[1][1]{\TFL{g}\sb{#1}}
\newcommand\ii{\mathrm{i}}

\newcommand\crt{j}

\newcommand\lp{\ell\sp\star}

\newcommand\pback{p\sb{\text{back}}}
\newcommand\proba{\mathbb{P}}
\newcommand\porig{\proba\sb{\text{origin}}}
\newcommand\pesc{\proba\sb{\text{escape}}}
\newcommand\poisson[1]{\mathcal{P}\sb{#1}}
\newcommand\grt{p}

\newcommand\meantime{\mathbb{T}}
\newcommand\torig[1][]{\meantime\sp{#1}\sb{\text{origin}}}
\newcommand\texit[1][]{\meantime\sp{#1}\sb{\text{exit}}}
\newcommand\gev[1][1]{u\sb{#1}}
\newcommand\TLgev[1][1]{\TL{u}\sb{#1}}

\newcommand\mx{\ominus}
\newcommand\px{\oplus}
\usepackage{relsize}
\newcommand\pmrx[3]%
  {\tikz[anchor=south,baseline=-#1]{
         \def\r{#2}
         \draw[line width=#3] (0,0) circle (\r); 
         \draw[line width=#3] (-0.968*\r,0.25*\r) -- (0.968*\r,0.25*\r);
         \draw[line width=#3] (0,-0.5*\r) -- (0,\r);
         \draw[line width=#3] (-0.866*\r,-0.5*\r) -- (0.866*\r,-0.5*\r);}}
\newcommand\pmx{\mathchoice
  {\pmrx{3pt}{0.12}{0.3}}
  {\pmrx{3pt}{0.12}{0.3}}
  {\pmrx{2.2pt}{0.09}{0.38}}
  {\pmrx{1.65pt}{0.05}{0.25}}
}

\newcommand\sign{\mathrm{sign}}
\newcommand\moy[2][]{\left\langle{#2}\right\rangle\sb{#1}}
\newcommand\TF[1]{\tilde{#1}}
\newcommand\TL[1]{\hat{#1}}
\newcommand\TFL[1]{\TL{\TF{#1}}}
\newcommand\TZ[1]{\overline{#1}}

\newcommand\derp[3][]{\partial\sp{#1}\sb{#3}{#2}}
\newcommand\convolution[1][]{\underset{#1}\otimes}
\newcommand\tx[1][x]{\sqrt{t\sp2-{#1}\sp2}}

\newcommand\cbal[1][+]{\ifx#1+{\dirac(ct-x)\e\sp{-ct/\lp\sb+}}
                        \else{\dirac(ct+x)\e\sp{-ct/\lp\sb-}}\fi}
\newcommand\bal[1][+]{\ifx#1+{\dirac(t-x)\e\sp{-t/\lp\sb+}}
                      \else\ifx#1-{\dirac(t+x)\e\sp{-t/\lp\sb-}}
                           \else\ifx#1\pm{\dirac(t\mp x)\e\sp{-t/\lp\sb\pm}}
                                \else{\dirac(t\pm x)\e\sp{-t/\lp\sb\mp}}
                                \fi
                           \fi
                      \fi}

\title{The one-dimensional asymmetric persistent random walk}
\author{Vincent \sc Rossetto
\footnote{\texttt{vincent.rossetto@grenoble.cnrs.fr}}\\
\small Université Grenoble Alpes and CNRS / LPMMC, Grenoble, France}
\date{}

\bibliographystylenote{iopart-num}

\begin{document}
\DeclareFontShape{T1}{EBGaramond-LF}{bx}{n}{ <-> ssub * cmr/bx/n }{}
\maketitle

\begin{abstract}
Persistent random walks are intermediate transport processes 
between a uniform rectilinear motion and a Brownian motion. 
They are formed by successive steps of 
random finite lengths and directions travelled at a fixed speed.
The isotropic and symmetric one-di\-men\-sio\-nal 
persistent random walk is governed by the telegrapher's equation,
also called hyperbolic heat conduction equation.
These equations have been designed to resolve the paradox of the
infinite speed in the heat and diffusion equations.
The finiteness of both the speed and the correlation length
leads to several classes of random walks: Persistent random
walk in one dimension can display anomalies that cannot arise
for Brownian motion such as anisotropy and asymmetries.
In this work we focus on the case where the mean free path is anisotropic,
the only anomaly leading to a physics that is different from the 
telegrapher's case. We derive exact expression of its Green's function, 
for its scattering statistics and distribution of first-passage time at 
the origin. The phenomenology of the latter shows a transition 
for quantities like the escape probability and the residence time. 
\footnote{
The first version of this article \citenote{rossetto2018}
contained typos, as mentioned in the work of 
Giona {\it et al.} \citenote{giona2019a}, that have
been corrected in this version.}
\end{abstract}

\section{Introduction}
Since it was introduced in Physics by Smoluchowski, Einstein and Langevin, 
Brownian motion has been studied and used in an ever growing range of
purposes to describe randomness of various kinds.
The probability distribution function
of Brownian motion is the same as the Green's function of the heat 
equation and the diffusion equation. However it was soon remarked that
these equations are not compatible with causality, because they do not
bound the instantaneous speed of motion. Brownian motion is therefore
only a limit of a more complex motion, observed at large time and 
length scales. At the atomic scale, the movement of the walkers is not
Brownian anymore, it is a sequence of rectilinear motions interspersed by
collisions. Such a movement is a random walk with steps of finite lengths
and finite time steps. Calling $\ell$ the average step length
--- the \emph{mean free path} --- and $\tau$ the
average time between two collisions, 
the speed is $c=\ell/\tau$. Brownian motion corresponds to 
the limit $\ell\to0$ and $\tau\to0$ when $\ell^2/\tau=D$ is kept constant.
Under this limit, the speed $\ell/\tau$ must diverge.

It is widely known that the spatial probability distribution of presence
of the Brownian motion is a Gaussian distribution. The Fokker-Planck
equation of Brownian motion is the diffusion equation and is
the same as the heat equation. Although these equations are
used in many physical models, they do not fulfill the requirement
of causality that is imposed to any physical equation. 
The absence of causality and the possible corrections to 
restore it in the heat equation have been discussed by 
Cattaneo~\cite{cattaneo1948} and Vernotte~\cite{vernotte1958}.
They independently proposed a microscopical theory yielding 
a supplementary term of second order time derivative
in the equation. The obtained equation is the so-called 
telegrapher's equation, first derived by Thomson
when he was studying the electric transport in the first 
transatlantic cable~\cite{thomson1855}.
The theory of heat waves has widely developped. For a complete 
review until 1989, refer to the exhaustive work of 
Joseph and Preziosi~\cite{joseph1989}. 
Other works on the diffusion equation, that suggested to add an
inertial term, obtained the same equation~\cite{brinkman1956,sack1956}.

In 1951, Goldstein showed that the telegrapher's equation is 
the Fokker-Planck equation of a persistent random walk
with a fixed time step~\cite{goldstein1951}. Persistent random walks were
introduced by Fürth~\cite{furth1920} and Taylor~\cite{taylor1922}.
In higher dimensions, another approach was proposed by 
Domb and Fisher~\cite{domb1958}.
In 1965, Montroll and Weiss generalized the random walks
to a continuous time process, with random time 
steps~\cite{montroll1965}. Masoliver, Lindenberg and Weiss 
introduced the persistent random walk with continuous time steps%
~\cite{masoliver1989}. 

The telegrapher's equation appears to be the one-dimensional 
version of the radiative transfer equation.
In scattering medium, waves propagate according to the radiative transfer
equation \cite{chandrasekhar}. In this equation the
direction of the partial waves scattered by an impurity is correlated
to the direction of the incoming wave. Consequently, the direction
of propagation has a correlation length called the \emph{persistence 
length} or the \emph{transport mean free path}~$\lp$. In one dimension,
this length can be different for walkers moving to the
right (direction denoted as $\px$) and walkers moving to the left
($\mx$). Such an
anisotropy can arise from the scattering mechanism itself, from internal
mechanisms or from external forces. It corresponds to the case where
the rate of change from $\px$ into $\mx$ is different from the
rate of change from $\mx$ into $\px$. 
Asymmetric persistent random
walks have been used to study the instability dynamics of microtubules%
~\cite{bicout1997}. In microtubule, the $\px$ and $\mx$ directions correspond
to growth or shrinkage of the one end of the microtubule. As these are 
different chemical processes, their characteristics speeds and
persistence lengths are different.
Persistent random walks have also been considered in the 
context of relativistic Brownian motion~\cite{dunkel2009}.

A random walker can reach, after some time, 
the boundary of its propagation domain, where it will be reflected
or absorbed, for instance by a detection device.
The location on the boundary and the time at which it is reached  
are relevant quantities to investigate the propagation properties. 
First passage problems are ubiquitous in physical sciences and have
been widely studied~\cite{redner,weiss1994}. Their
properties are mostly studied for Brownian motion in several situations
in two and even in higher dimensions \cite{masoliver1993some}.
The statistics of the first passage times of the persistent random walk
have been studied by Masoliver, Porr\`a and 
Weiss~\cite{masoliver1992solutions}.

Most of the cited studies of the one-dimensional persistent random walk
have been performed in the case where the mean free path is isotropic.
In this article, we start by showing that mean free path anisotropy is
actually the only candidate for new physics within all possible 
transport and scattering anomalies in one dimension. The next sections
of this work are dedicated to the derivation and solution of the asymmetric
equation. We show that the number of scattering events follows a distribution
obtained using the same functions. In the last section, we study the
first-passage time distribution at the origin and show that it posseses
a very intuitive phenomenology and transitions around the symmetry point.

\section{The asymmetric one-dimensional persistent random walks}
In this section we discuss the properties of one-dimensional
random walks and their asymmetries. The one-dimensional space
is indexed by~$x\in\mathbf{R}$, a real number.
In one dimension only two directions of propagation are possible,
$\px$ and $\mx$,
which makes one dimensional random walks particular. 

We consider a
random walker moving along a one dimensional space containing scatterers. 
The positions of the scatterers are not fixed, they uniformly distributed
along space and time. (This has the advantage to yield 
a time-invariant model.)
The 
walker moves at a constant speed $c$ as long as it does not encounter any
scatterer. When it meets a scatterer, it has a certain
probability to change its directions. If it does not so, there is no observable
sign that a scatterer was actually met. Contrary to the Brownian walker,
the persistent walker has a definite velocity at almost all times. We can
therefore count the time it spends moving in the $\px$ direction
during the movement. We will denote this time by $t^+$. Similarly,
the time spent moving in the $\mx$ direction is denoted by $t^-$. 
The total time of displacement is of course $t=t^++t^-$. If the
displacement during this time is $x$, we have $x=c(t^+-t^-)$,
therefore we easily obtain $t^\pm=\frac12(t\pm x/c)$. 

\subsubsection*{Scattering anisotropy}
In most of the studies of the persistent random walk,
scattering is considered as instantaneous, isotropic and non-absorbing.
We denote $\pback$ the probability that the walker changes its direction 
after a scattering event.
The walker moves into one direction along an average distance of
$\lp=\ell/\pback$ (Note that if $\pback=0$, there is no
scattering and the motion is rectilinear and uniform.) The
length $\lp$ is called \emph{the transport mean free path}. 
In one dimension all anisotropic
persistent random walks sharing the same transport mean free path
are fully equivalent. The most studied case is the isotropic scattering case 
where $\pback=1/2$ and $\lp=2\ell$ \cite{weiss1981}. This case
corresponds to the situation where the random walker is isotropically 
redirected as if it were a photon scattered in a cloud of point-like 
scatterers. Another
interesting situation is the backward scattering case, 
where $\pback=1$, because the two mean free paths are equal.
The backward scattering case is more intuitive and can be used to
interpret more easily the distribution of probability of the position
and direction of the random walker. In this work, we consider,
without loss of generality, only the scattering mean free path~$\lp$
such that we will consider the situation $\pback=1$. 

\subsubsection*{Transport asymmetries}
We speak of asymmetry in the case where a physical constant 
for the walkers travelling in the $\px$ direction  has a different value
for the walkers travelling in the $\mx$ direction. 
Since the physical constants describing the persistent random
walk are the speed~$c$, the transport mean free path~$\lp$ and
the absorption~$a$, we
introduce the speeds~$c^+$ and $c^-$, the transport mean free
paths $\lp_+$ and $\lp_-$ and the absorptions $a^+$ and $a^-$,
for the walkers moving in the directions $\px$ or $\mx$
according to the sign.

A speed asymmetry can account for different polymerization and
depolymerization rates in the case of microtubule dynamics~\cite{bicout1997}.
The Galilean change of variable $x'=x-vt$, with $v=({c^+-c^-})/2$,
restores the speed symmetry
in the translation frame $(x',\,t)$ in which the speed is $c=(c^++c^-)/2$. 
An absorption asymmetry damps the whole solution
by a factor $\exp(-a^+t^+-a^-t^-)$ where $t^\pm$
is the time spent in the direction $\pmx$.
Speed and absorption asymmetries can therefore be disposed of easily,
all results obtained in the case where $c^+=c^-$ and $a^+=a^-=0$
can be extended to the general case by using a change of referential
and by adding a damping factor.

We will focus our discussion on the mean free path asymmetry that 
yields non trivial features.
We use the two transport mean free paths~$\lp_+$ and $\lp_-$ 
to define the average transport mean free path~$\lp$, 
the attenuation rate~$\mu$ and the 
the asymmetric wavenumber $\kappa$ as
\begin{equation}
\lp=\sqrt{\lp_+\lp_-}\;
,\;\;
\mu=\frac{c}{2}\left(\frac{1}{\lp_+}+\frac{1}{\lp_-}\right)
\;\text{and}\;\;
\kappa=\frac{1}{2}\left(\frac{1}{\lp_-}-\frac{1}{\lp_+}\right).
\label{definitions}
\end{equation}
They are related by $\mu^2/c^2-\kappa^2=1/(\lp)^2$.
The sign of $\kappa$ characterizes the direction in which the
transport mean free path is larger. If $\kappa>0$, we have $\lp_+>\lp_-$
such that the average distance travelled by the walker between two scattering 
events is larger in the $\px$ direction than in the $\mx$. As a result,
the mean position of the walker is drifting in the $\px$ direction. 

In the next section, we derive the Fokker-Planck equation governing
the probability density function of the position, that we call the
asymmetrie telegrapher's equation (ATE).

\section{The asymmetric telegrapher's equation}
\label{ATE}
We denote by 
$p^+(x,\,t)$ and~$p^-(x,\,t)$ the spatial probability densities
at time~$t$ of walkers moving in the direction~$\px$ and~$\mx$ respectively.
We have the following coupled master equations
\begin{equation}
\left\{\begin{array}{rcl}
  p^+(x,\,t+\Delta t)&=&
    \left(1-\frac{\Delta x}{\lp_+}\right)p^+(x-\Delta x,\,t) 
                 + \frac{\Delta x}{\lp_-}p^-(x+\Delta x,\,t), \\
  p^-(x,\,t+\Delta t)&=&
    \left(1-\frac{\Delta x}{\lp_-}\right)p^-(x+\Delta x,\,t) 
                 + \frac{\Delta x}{\lp_+}p^+(x-\Delta x,\,t).
\end{array}\right.
\label{equation maitresse 1d}
\end{equation}
Let us expand to the first order in~$\Delta x$ and $\Delta t$,
and use the relation $\Delta x=c\Delta t$. We obtain
two coupled partial differential equations
for the total probability density~$p=p^++p^-$ and the
current density~$\crt=c(p^+-p^-)$:
\begin{equation}
  \derp p t=-\derp \crt x,\qquad
  \derp\crt t=-c^2\derp p x-2\mu\crt+2\kappa c^2\, p.
\label{eq pj}
\end{equation}
The first of these equations is the continuity equation that ensures 
the conservation of probability. In the case where there is absorption,
this equation is no longer satisfied. 
Eliminating~$j$ we find
\begin{equation}
\left(\derp[2]{}x-\frac1{c^2}\derp[2]{}t
-2\kappa\derp {} x -\frac{2\mu}{c^2} \derp {} t\right)p=0.
\label{telegrapher}
\end{equation}
Let us remark that in the symmetric case $\kappa=0$,
if~$\mu\to0$ ($\lp\to\infty$), we obtain a one-dimensional wave equation
whereas if~$c\to\infty$ while the value of~$c^2/\mu$ is kept constant, one
obtains a diffusion equation.
The equation~\eqref{telegrapher} is an interpolation between these two
equations. In the ``wave'' limit, disorder is lost and in the diffusive
limit, causality is lost. 
In the symmetric case $\kappa=0$, the equation~\eqref{telegrapher} 
is called the telegrapher's equation. 
Observe that, thanks to the continuity equation, $j$ is also solution
of this equation (with different initial conditions)
and consequently so are $p^+$ and $p^-$. 

For convenience, we use the length and time units such that
$\lp=1$, $c=1$. 
The dimension of $p^+$, $p^-$ and $\grt$ is an inverse length
$\left[\mathrm{L}^{-1}\right]$, the
dimension of~$\crt$ is an inverse time $\left[\mathrm{T}^{-1}\right]$. 
The symmetric case is retrieved using $\mu=1$, $\kappa=0$.

\section{Solutions of the asymmetric telegrapher's equation}
\label{solATE}
\subsection{General solutions}
Let us solve the equation \eqref{telegrapher}
using both the Fourier transform in 
space and the Laplace transform in 
time. We denote by $\TL f(x,\,s)$ the Laplace
transform of $f(x,\,t)$ and by $\TF f(k,\,t)$ its Fourier transform.
The double transform is denoted by~$\TFL f(k,\,s)$.
Equation~\eqref{telegrapher} becomes
\begin{equation}
\TFL f(k,\,s)=\frac{(s+2\mu)\TF f(k,\,0)+\derp{\TF f}t(k,\,0)}
             {s^2+2\mu s+k^2-2\ii k\kappa}
\label{FL Q}
\end{equation}
I introduce the elementary function~$\gscat$ defined by
\begin{equation}
\TFLgscat(k,\,s)=\frac1{(s+\mu)^2+(k-\ii \kappa)^2-1}.
\label{def Gscat}
\end{equation}
The function~$\gscat$ is ubiquitous in the statistical theory
of the persistent random walks. The equation~\eqref{def Gscat}
shows that it is a non-persistent multiple scattering Green's function,
but $\gscat$ is \emph{not} the Green's function of the telegrapher's
equation~\eqref{telegrapher}.
Performing the inverse Fourier transform we get 
\begin{equation}
\TLgscat(x,\,s)=
    \frac12\frac{\e^{\kappa x-|x|
    \sqrt{(s+\mu)^2-1}}}{\sqrt{(s+\mu)^2-1}}
\label{TL Gscat}
\end{equation}
and the inverse Laplace
transform is found in Ref.~\cite[29.3.93]{abramowitzstegun}
for~$\TLgscat(x,\,s-\mu)$, which 
provides~$\gscat(x,\,t)\e^{\mu t}$. We define more generally~$\gscat[z]$ as
\begin{equation}
\gscat[z](x,\,t)=z\frac{\e^{\kappa x-\mu t}}{2}
  I_0\left(z\tx\right)\Theta(t-|x|)
\label{def g}
\end{equation}
and use $\gscat[1]$ in this section. The variable~$z$ will be used in the
next section.
We also introduce the function $\gev[z]$
\begin{equation}
\gev[z](x,\,t)=z\frac{\e^{\kappa x-\mu t}}2\frac{I_1\left(z\tx\right)}\tx
 \Theta(t-|x|)
\label{def gev}
\end{equation}
that appears in many developped expressions. Both~$\gscat$ and $\gev$ have a
probabilistic interpretation that we discuss in the Section~\ref{proba}.
Some mathematical relations between $\gscat[z]$ and $\gev[z]$ are given
in the appendix.

The solution to~\eqref{telegrapher} 
in the~$(x,\,s)$ domain with the initial conditions
$f(x,\,0)=v(x)$ and $\derp ft(x,\,0)=w(x)$
is obtained by spatial convolution from the expression~\eqref{FL Q}
as 
\begin{equation}
\TFL f(x,\,s)=(s+2\mu)\TLgscat(x,\,s)\convolution[x]v(x)
+\TLgscat(x,s)\convolution[x]w(x).
\label{TL sol ATE}
\end{equation}
In the time domain, using~\eqref{u0t}, it translates into
\begin{equation}
   \derp \gscat t(x,\,t)\convolution[x]v(x)+
   \gscat(x,t)\convolution[x]\left(2\mu v(x)+w(x)\right).
\label{sol ATE}
\end{equation}

\subsection{The Green's function of the ATE in a infinite domain}
We denote by $\grt(x,\,t\,|\,x_0,\,t_0)$ the solution of
the ATE with the initial position in $x_0$ at $t=t_0$. 
The Green's function of the ATE~\eqref{telegrapher} is 
$\grt(x,\,t\,|\,0,\,0)$. If there is no ambiguity, we write
it simply as $\grt(x,t)$. The direction of propagation comes
as an extra argument in the variables or in the initial conditions.
The Green's function with imposed initial velocity has initial
conditions 
$\grt(x,\,0\,|\,\pmx)=\dirac(x)$ 
and~$\crt(x,\,0\,|\,\pmx)=\pm\dirac(x)$.
To use these initial conditions in equation~\eqref{sol ATE}
we compute
$\derp\grt t(x,\,0\,|\,\pmx)=-\derp\crt x(x,\,0\,|\,\pmx)=\mp\dirac(x)$
which immediately gives the solution expressed by means of~$\gscat$
\begin{equation}
\grt(x,t|\pmx)= \derp\gscat t(x,t) +2\mu\gscat(x,t) \mp\derp\gscat x(x,t).
\label{P0}
\end{equation}
Using the relations~\eqref{u0t} and~\eqref{u0x}, 
we readily obtain the expression of $\grt$ as
\begin{equation}
 \grt(x,\,t|\,\pmx)=
  \frac1{\lp_\pm}\gscat(x,\,t) +(t\pm x) \gev(x,\,t) +\bal[\pm].
\label{hemmer}
\end{equation}
This result is made of three terms. The scattering contributions~$\gscat$
and $\gev$ contain a step function~$\Theta$ that ensures causality.
The term with a Dirac distribution~$\dirac$ is the contribution of the walkers
that have not been scattered at all. 
This solution was obtained in the symmetric
case~$\lp_+=\lp_-$ for the initial condition where half the walkers are~$\px$
and half are $\mx$ by Goldstein~\cite{goldstein1951}, 
Morse and Feschbach~\cite[p. 865]{morsefeschbach1} and
Hemmer~\cite{hemmer1961}.
It was later interpreted as the solution
of the symmetric radiative transfer in one dimension 
by Paasschens~\cite{paasschens1997}.
Proceeding with the same method, using $v(x)=\pm\dirac(x)$ 
and $w(x)=-\ddirac(x)\mp2\mu\dirac(x)+2\kappa\dirac(x)$ 
in \eqref{sol ATE}, we find
\begin{equation}
\crt(x,\,t\,|\,\pmx)=\pm\derp\gscat t(x,\,t)
+ 2\kappa\gscat(x,t)-\derp\gscat x(x,t),
\label{hemmerjg}
\end{equation}
or, in terms of $g$ and $u$
\begin{equation}
\crt(x,\,t\,|\,\pmx)=\pm\left[-\frac{1}{\lp_\pm}\gscat(x,\,t)
  +(t\pm x)\gev(x,t)+\bal[\pm]\right].
\label{hemmerj}
\end{equation}
We observe that $\crt$ has the same terms as~$\grt$
with different signs. This fact will find an explanation in the
Section~\ref{proba}. 

\begin{figure}
\begin{center}
\includegraphics[width=0.47\textwidth]{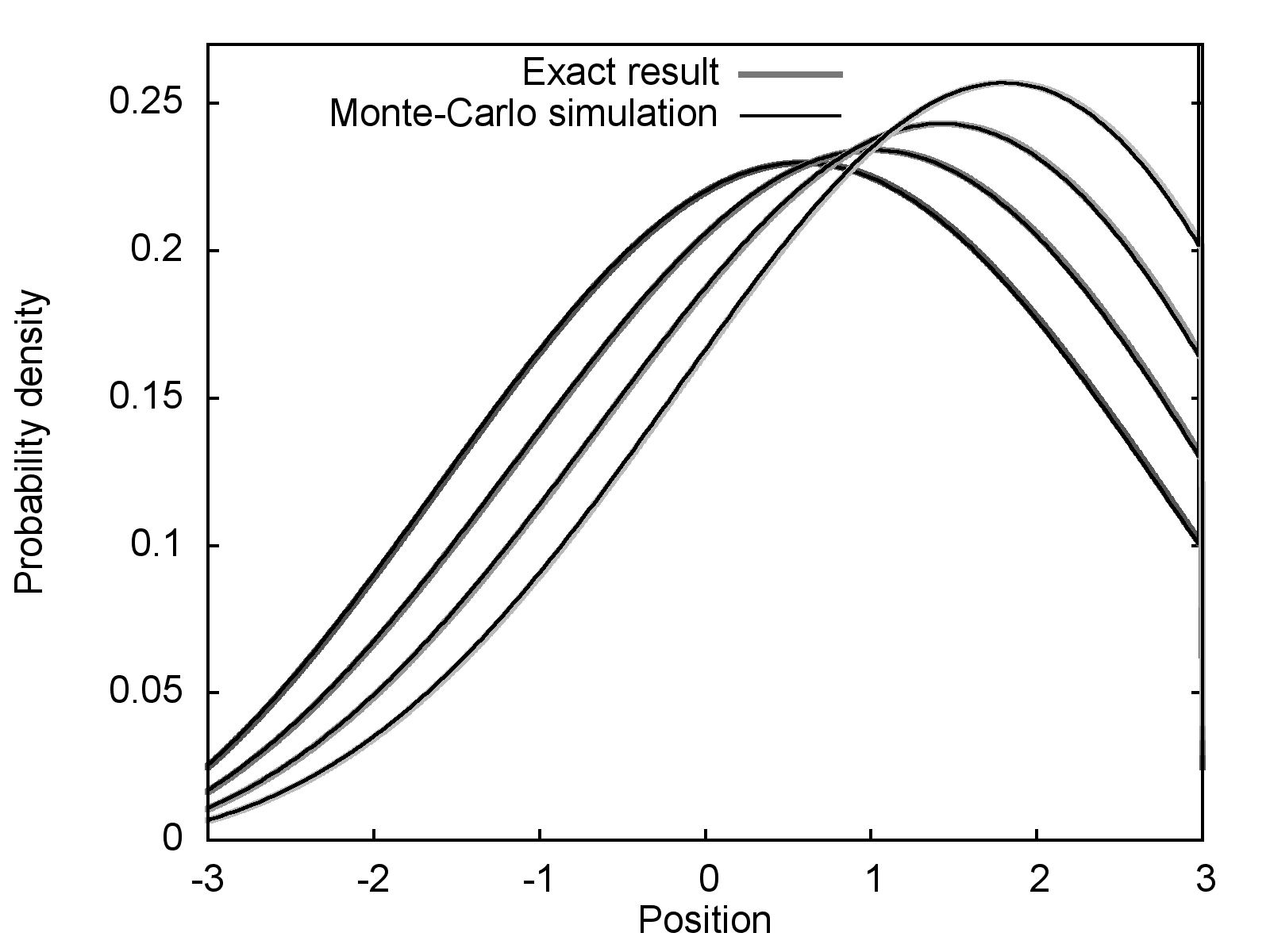}
\includegraphics[width=0.47\textwidth]{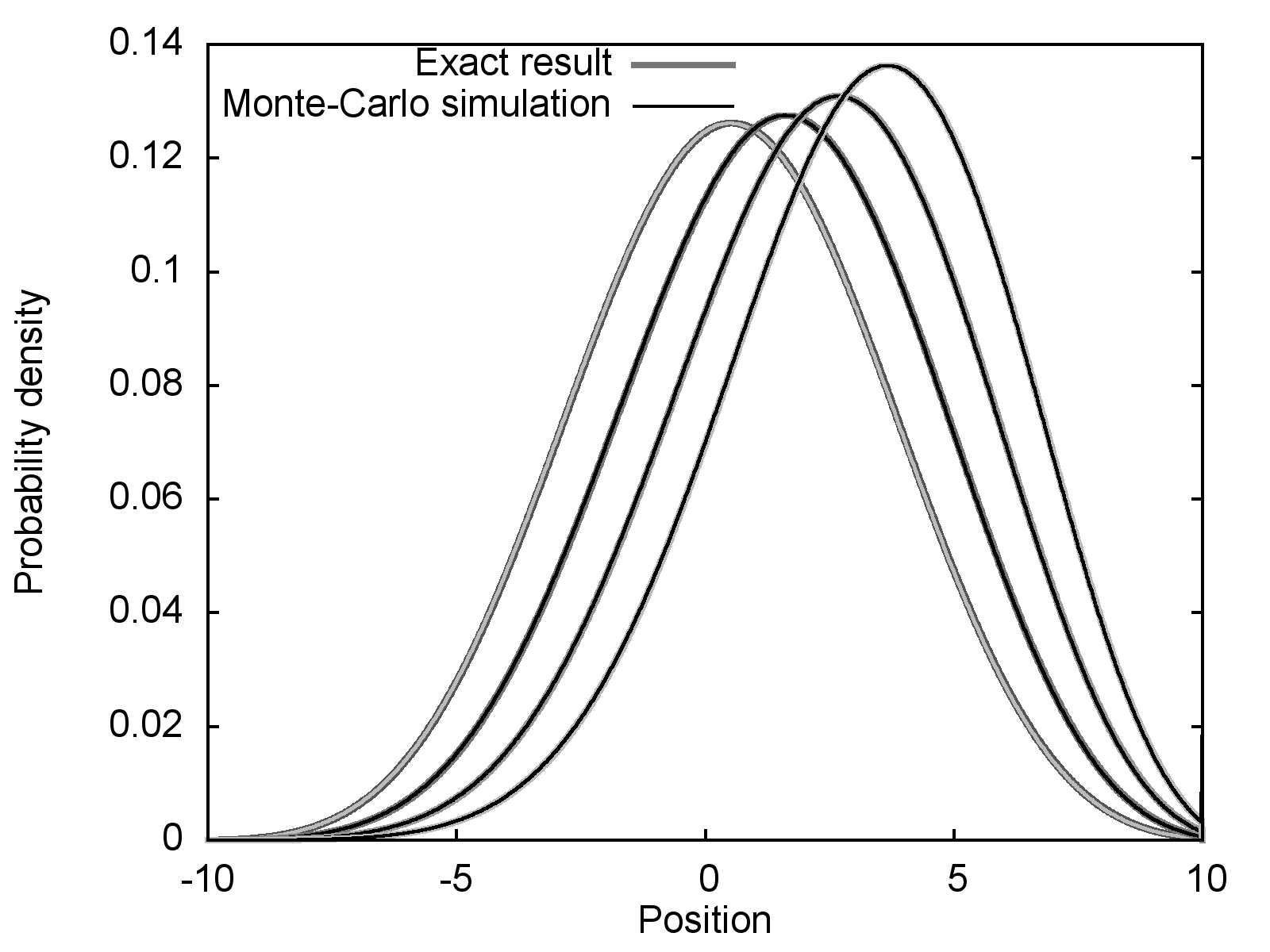}
\caption{\label{figure p}Spatial distribution of the persistent
random walker after a time $t=3$ (Left) and $t=10$ (Right) for
different values of the asymmetry parameter~$\kappa$. The unbiased
distribution ($\kappa=0$) reaches its maximum at $x=0$. The other
distributions have been obtained for $\kappa=0.1$, $\kappa=0.2$ and
$\kappa=0.3$. The exact result of Equation~\eqref{hemmer} is displayed with
a thick line. The superimposed thin line has been obtained using a
Monte-Carlo simulation of the persistent random walk. The random walk
starts in the direction~$\px$. Each step length is randomly generated
using an exponential distribution $\mathrm p(x)=\ell^{-1}\e^{-x/\ell}$ 
where $\ell=\lp_+$ or $\lp_-$ according to the direction of propagation.
After moving along one step, the direction is changed into its opposite
(corresponding to the case $\pback=1$). The statistics have been performed
using $10^{10}$ independent realizations of the random walk.}
\end{center}
\end{figure}

\subsection{Direction of propagation}
Let us consider the walkers starting in the~$\px$ state 
and compute~$p^\pm=(\grt\pm\crt)/2$ using the
expressions~\eqref{hemmer} and \eqref{hemmerj}. We obtain 
\begin{align}
p^+(x,\,t\,|\,\px)&=(t+x)\gev(x,\,t)+\bal&
\text{and}&&p^-(x,\,t\,|\,\px)&=\frac{1}{\lp_+}\gscat(x,\,t).
\label{p+1}
\end{align}
For the walkers starting in the~$\mx$ state, we find
similar results
\begin{align}
p^-(x,\,t\,|\,\mx)&=(t-x)\gev(x,\,t)+\bal[-]&
\text{and}&&p^+(x,\,t\,|\,\mx)&=\frac{1}{\lp_-}\gscat(x,\,t).
\label{p-1}
\end{align} 
Note that the above expressions are not normalized, but that normalization
is made straightforward by noticing that the spatial integral of the 
function~$\gscat$ is equal to $(1-\e^{-2\mu t})/(2\mu)$ and that
$p$ is normalized. We observe then that when the time is growing, the
fraction of walkers moving in the direction~$\pmx$ is 
$\frac{\mu\pm\kappa}{2\mu}$, independently of the initial direction.

The interpretation of these expressions is that in the
equations~\eqref{hemmer} and \eqref{hemmerj}, the terms containing~$\gev$
account for the walkers that are at time~$t$ travelling in the same direction
as at $t=0$ and have been scattered at least once.
The terms containing~$\gscat$ account for 
walkers travelling in the opposite direction as at $t=0$.
This remark is refined in the next section, using a more detailed
statistics of the scattering events and their time distributions.

\section{Counting statistics of scattering events}
\label{proba}
In the previous, we have derived the solution of the ATE using
analytical methods. The results known for the symmetrice telegrapher's
equation have been retrieved in the symmetric case. We ought to extend
to the asymmetric case the most relevant contribution concerning statistics 
concerning the counting statistics of scattering event \cite{foong1994},
the first-passage times~\cite{masoliver1992solutions,masoliver1992erratum} 
and maximum displacement~\cite{masoliver1993maximum}.

We recall some elementary facts concerning the exponential 
probability distribution. Let us consider $n$ independent events during
a time $t$, such that the waiting time is distributed
exponentially with probability distribution $p(t)=\frac1\tau\exp(-t/\tau)$. 
If the total duration~$t$ is fixed, the number
of events is distributed according to a Poisson law:
\begin{equation}
\poisson{}(n\,|\,t)=\frac{(t/\tau)^n}{n!}\e^{-t/\tau}.
\label{poisson}
\end{equation}
We rephrase this law by stating that the duration~$t$ is split into
$n+1$ time intervals with probability $\poisson{}(n)$.  
If the number of events~$n$ is fixed then the total
duration~$t$ is distributed according to a gamma law~$\Gamma(n,\frac1\tau)$
\begin{equation}
\Gamma(t\,|\,n)=\frac{1}{\tau}\frac{(t/\tau)^{n-1}}{(n-1)!}\e^{-t/\tau}.
\label{gamma}
\end{equation}

We shall finally remark that after a time $t$ 
the persistent random walk only reaches
positions $|x|\leq t$ and that the case of equality
is reached only for trajectories with no scattering events. These
trajectories are trivial and are treated separately. As a consequence, 
we consider in this section that $|x|<t$. Using the prescription
the Dirac-delta functions disappear and all the functions are analytical.

\subsection{Probability of the number of scattering events}
We consider now the random walk starting at $(0,0)$ in the $\px$
direction and ending in $(x,t)$ in the $\px$ direction. Let us
call $2n$ the number of scattering events (equal in our case to the
number of changes of direction because $\lp=\ell$). 
The probability density of
this walk is $p^+(x,t\,|\,\px)$, given by the Equation~\eqref{p+1}. 
Let us consider the time interval $t^+$. The probability
that $n$ scattering event happen during this time is $\poisson+(n\,|\,t^+)$.
If there are $n$ scattering events in the $t^+$ interval,
there must be the same number $n$ events in the $t^-$ interval
so the distance $t^-$ travelled in the $\mx$ direction is distributed
according to $\Gamma(t^-\,|\,n)$. 
The probability of having $2n$ scattering events is obtained from the
Bayes formula
\begin{equation}
N(2n\,|\,x,\,t\,;\,\px)=
 \frac{\poisson+(n\,|\,t^+)\Gamma_-\left(t^-\,|\,n\right)\dd t^-}
      {p^+(x,\,t\,|\,\px)\dd x}
\end{equation}
where $\poisson\pm$ is the Poisson law of rate $1/\lp_\pm$
and $\Gamma_\pm$ is the gamma law of rate $1/\lp_\pm$.
Since we are in the case where $n>0$, the normalisation of $N$ gives
\begin{equation}
\frac1{p^+(x,\,t\,|\,\px)}\,\frac{\dd t^-}{\dd x\;}\;
   \sum_{n>0}\poisson+(n\,|\,t^+)\Gamma_-(t^-\,|\,n)=1
\label{sommep++}
\end{equation}
or equivalently, using $\dd t^-/\dd x=1/2$, 
\begin{equation*}
\begin{split}
p^+(x,\,t\,|\,\px)&=\frac12\;
  \sum_{n>0} \frac1{n!}\left(\frac{t^+}{\lp_+}\right)^n
                       \e^{-t^+/\lp_+}\frac1{\lp_-}
       \frac1{(n-1)!}\left(\frac{t^-}{\lp_-}\right)^{n-1}
                       \e^{-t^-/\lp_-}\\
&=\frac{t+x}2\;
  \frac{I_1(\sqrt{t^2-x^2})}{\sqrt{t^2-x^2}}
  \e^{\kappa x-\mu t}
\end{split}\end{equation*}
where $\mu$ and $\kappa$ are defined as in Eq.~\eqref{definitions}. 
This computation is an independent demonstration of the equality
\begin{equation}
p^+(x,\,t\,|\,\px)=(t+x)\,\gev(x,\,t). 
\label{p+stat}
\end{equation}
It differs from the Equation~\eqref{p+1}, because the Dirac delta function
accounting for the case $n=0$ has been removed from the analysis.
With similar reasoning, we find that if the direction is $\mx$ at time~$t$,
the number of scattering events is $2n+1$, with $n\geq0$ 
distributed according to
$\poisson-(n\,|\,t^-)$ and the time~$t^+$ distributed according to
$\Gamma_+(t^+\,|\,n+1)$. With a similar summation as 
Equation \eqref{sommep++}, we
find $p^-(x,\,t\,|\,\px)=\frac1{\lp_+}g_1(x,\,t)$ as expected from 
Equations~\eqref{p+1}. Exchanging the signs, we find the results of
Equations~\eqref{p-1}. 

\subsection{Generating function of the scattering statistics}
Now that we have obtained the probability distribution of the number of
events, it is straightforward to introduce 
the generating function of the number of scattering events.
The generating
function of $N(n\,|\,\,x,\,t)$ is defined by
\begin{equation}
  \TZ N(z\,|\,x,\,t,\,\pmx)
  =\sum_{n=0}^\infty z^n N(n\,|\,x,\,t,\,\pmx)=
  \frac{\frac{1}{\lp_\pm}\gscat[z](x,t)+(t\pm x)\gev[z](x,t)}
       {\frac{1}{\lp_\pm}\gscat[1](x,t)+(t\pm x)\gev[1](x,t)}.
\label{generating sigma}
\end{equation}
Note that if $\pback<1$, it suffices to replace $z$ by $\frac{\pback
z}{1+\pback z-z}$ in this expression to get the counting statistics of
scattering events. From the expression~\eqref{generating sigma}, all the 
statistics of the number of scattering events
can be derived. In particular, the average number of scattering events
is $\moy{n(x,\,t,\,\pmx}={\TZ N}'(1\,|\,x,\,t,\,\pmx)$. 
The fluctuations of the number of scattering events are given by
\begin{equation*}
\big(\Delta n(x,\,t,\,\pmx)\big)^2
={\TZ N}''(1\,|\,x,\,t,\,\pmx)+\moy{n(x,\,t,\,\pmx)}-\moy{n(x,\,t,\,\pmx)}^2
\end{equation*}
It is worth noticing that the same method applies for the walkers
that $\px$ or $\mx$ at time $t$. The expressions of the averages 
and fluctuations of $n^\pm$ are given in the Appendix~\ref{stat n+/-}.
Notice also that if the scatterers absorb a fraction $a$ of the walkers,
the Green's function becomes 
$p(x,\,t,\,\pmx)=\TZ N(1-a\,|\,x,\,t\,\pmx)
  +\dirac(t\mp x)\exp(-t/\lp_\pm)$. 

\section{First-passage statistics at a fixed boundary}
\label{fp}
\subsection{General theory}
\begin{figure}
\begin{center}
\includegraphics[width=0.5\textwidth]{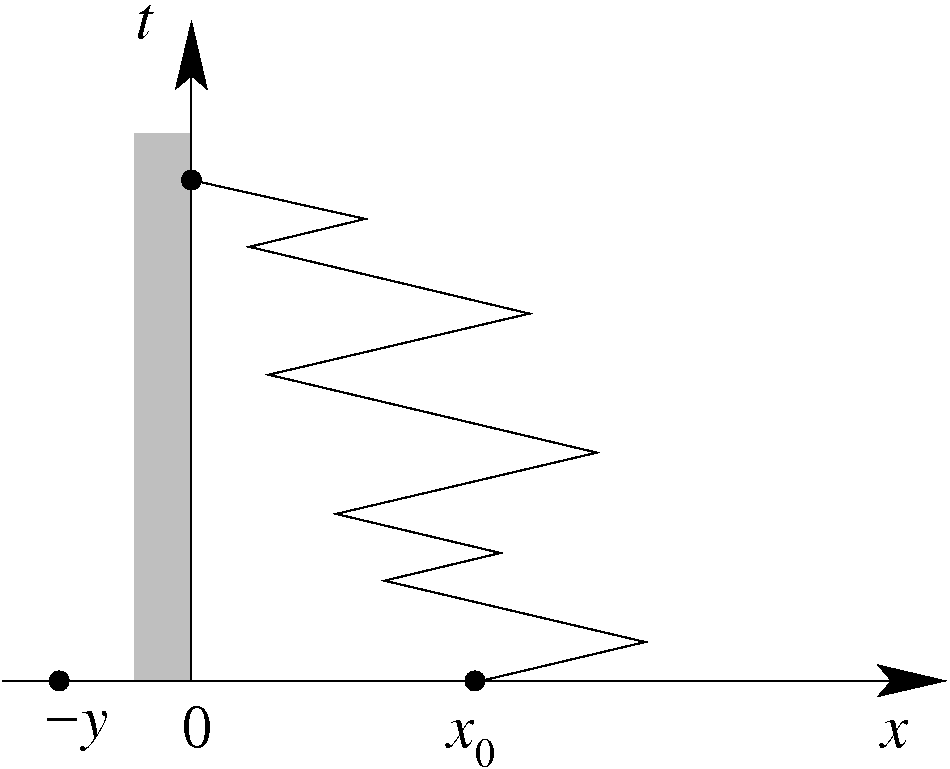}
\caption{\label{trap 1d} Persistent random walk with a trap at the origin.
The random walker starts in~$x=x_0$ and moves at constant speed~$c$. 
It changes its direction at random times (Poisson distributed, 
with mean~$\lp/c$). The process stops
when the random walker reaches the boundary at the origin. The time
at which this event occurs is called the first-passage time at the origin.
The point located at $-y$ (with $y>0$) is an intermediary used during
the computation of the first passage time probability distribution.}
\end{center}
\end{figure}
We consider now the first-passage problem at the origin while
the walk starts at~$x_0>0$. To obtain the time
distribution of the first arrival at the origin, 
we use Siegert's formula \cite{siegert1951} in the same way as 
Foong and Kanno did for the symmetric telegrapher's equation \cite{foong1994}. 
We denote by $r_1(t\,|\,x_0,\,\pmx)$ the time probability distribution of
the first passage at the origin of the random walker. Siegert's formula states
that, for any $y>0$ 
\begin{equation}
\TL r_1(s\,|\,x_0,\,\pmx)=
  \frac{\TL\grt(-x_0-y,\,s\,|\,\pmx)}{\TL\grt(-y,\,s\,|\,\mx)},
\label{siegert}
\end{equation}
where $\TL\grt$ is the Laplace transform of the Green's function~$\grt$
as given by Equation~\eqref{hemmer}. The first-passage time distribution
has the dimension of an inverse time $\left[\mathrm T^{-1}\right]$. 
It follows from the fact
that to reach the point $-y$ (see figure \ref{trap 1d}), the walker
must first reach the origin and start from this point a random walk with
initial velocity $-c$. We have readily
\begin{equation}
  \TL\grt(x,\,s\,|\,\pmx)=\frac{1}{2}
   \left(\frac{s+2\mu\mp\kappa}{\sqrt{(s+\mu)^2-1}} \pm\sign(x)\right)
  \e^{\kappa x-|x|\sqrt{(s+\mu)^2-1}}.
\end{equation}
In the $\mx$ case, the expression takes the simple form
$\TL r_1(s\,|\,x_0,\,\mx)=\e^{-\kappa x_0-x_0\sqrt{(s+\mu)^2-1}}$ such that
using the Equation~\eqref{U} we obtain
\begin{equation}
r_1(t\,|\,x_0,\,\mx)=2x_0\,\gev(-x_0,\,t)+\dirac(t-x_0)\e^{-x_0/\lp_-}.
\label{r-}
\end{equation} 
We remark that the result does not depend on $y$, as should be expected. 
We should also notice that the result is made of direct contribution
for the unscattered walkers and a term containing~$\gev$ which accounts
for an even number of scattering events. The counting statistics
of the number of scattering events before reaching the origin
is indeed obtained by replacing the functions in Siegert's formula 
\eqref{siegert} by the generating functions $\TZ P(\cdot,\,s,\,z\,|\pmx)$
The generating function
of the number of scattering events before reaching the origin 
is therefore $r_z/r_1$
with $r_z(t\,|\,x_0,\,\mx)=
2x_0\gev[z](-x_0,\,t\,|\,\mx)$. (The Dirac delta
function accounting for zero scattering events should be discarded 
in this expression for the same reasons as in the preceding section.)

\begin{figure}
\begin{center}
\includegraphics[width=0.47\textwidth]{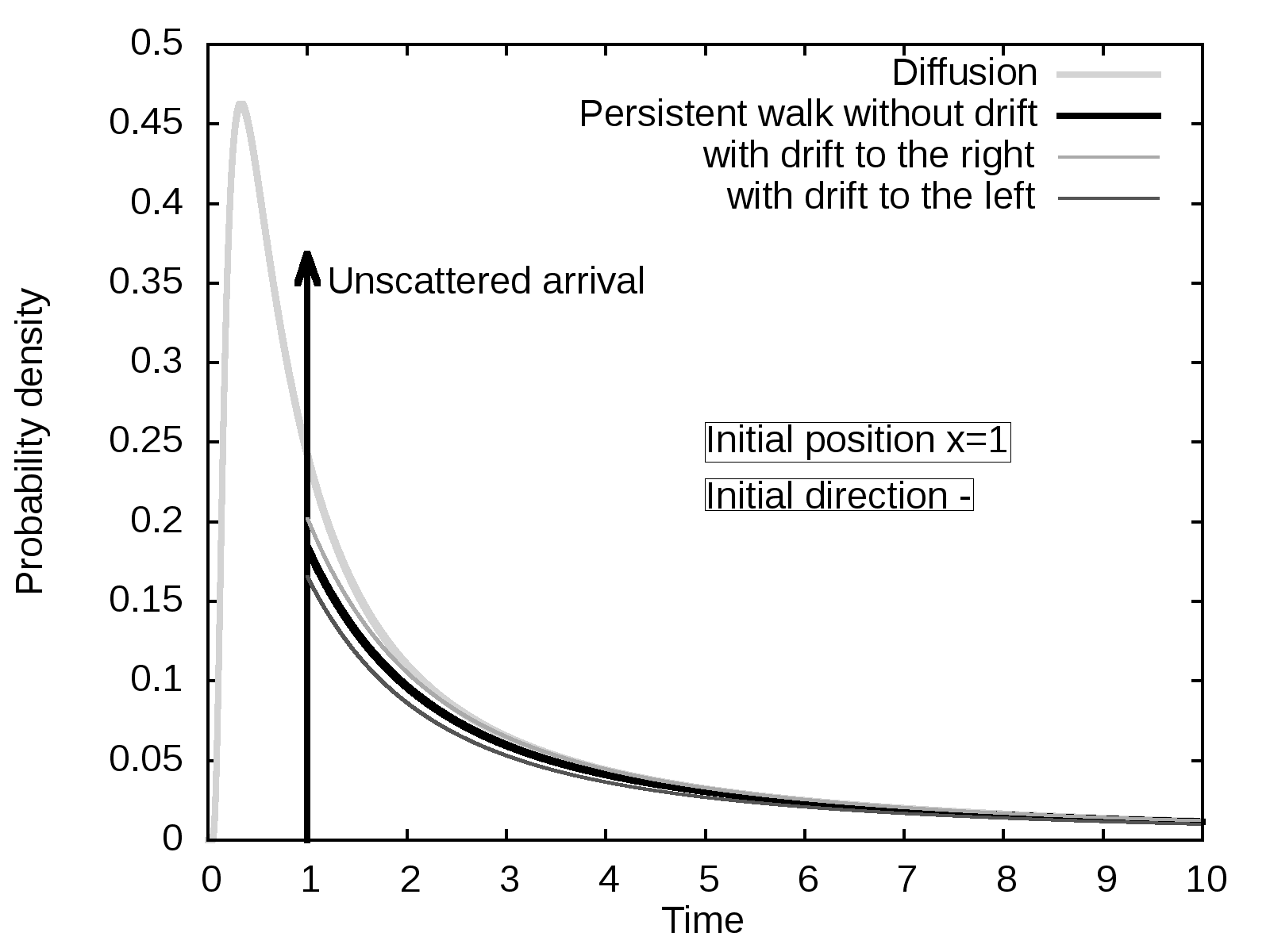}
\includegraphics[width=0.47\textwidth]{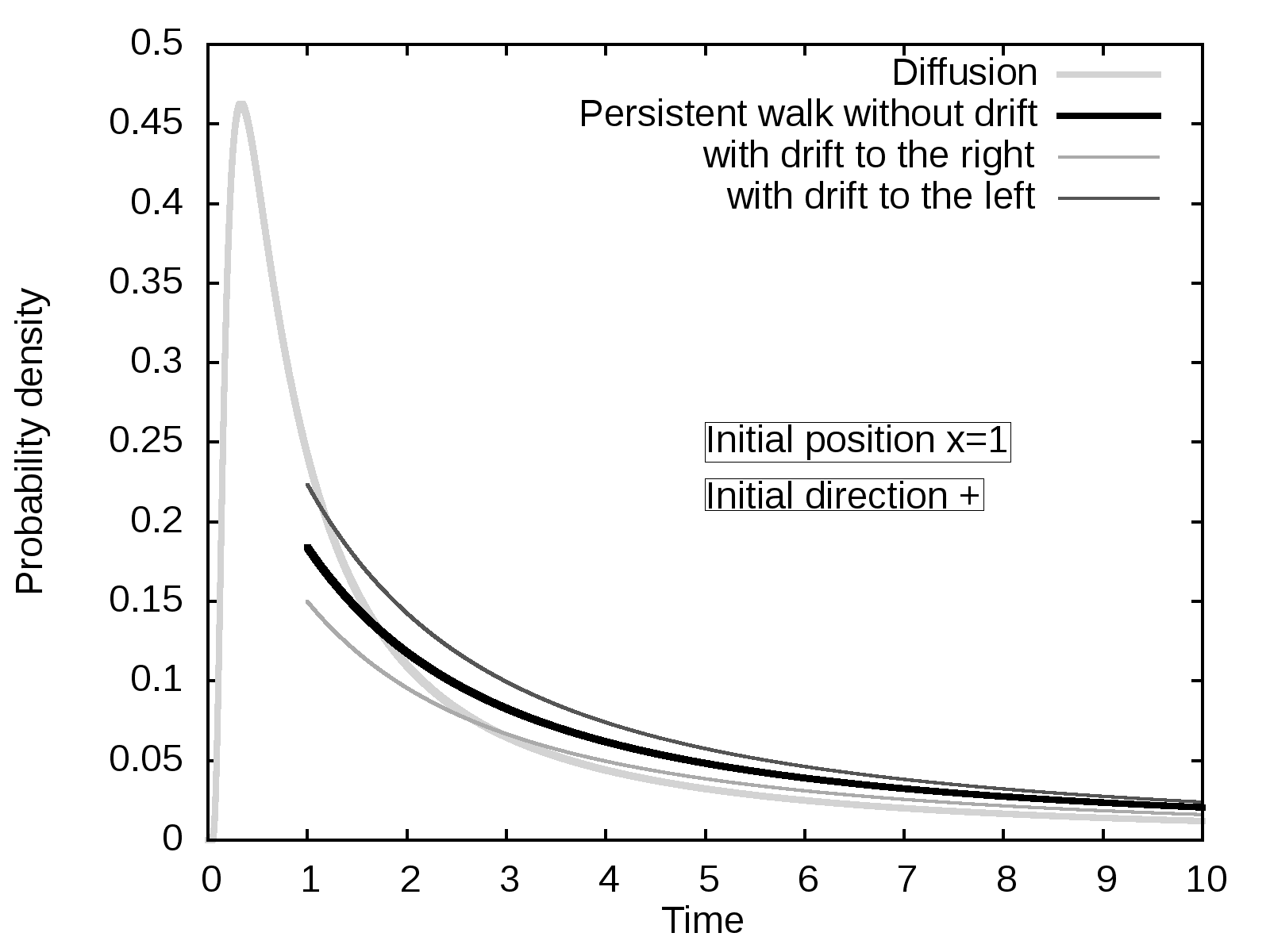}

\includegraphics[width=0.47\textwidth]{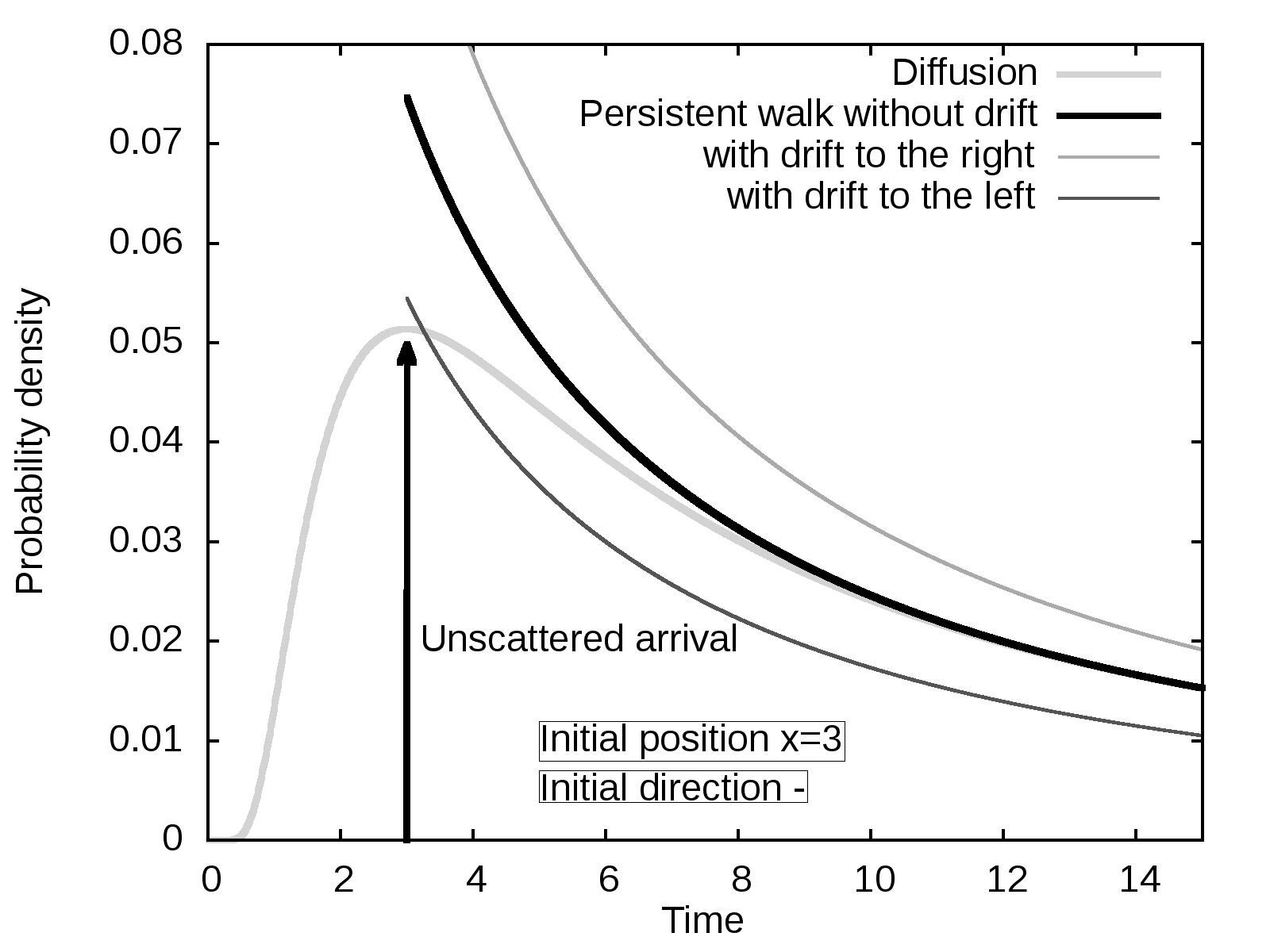}
\includegraphics[width=0.47\textwidth]{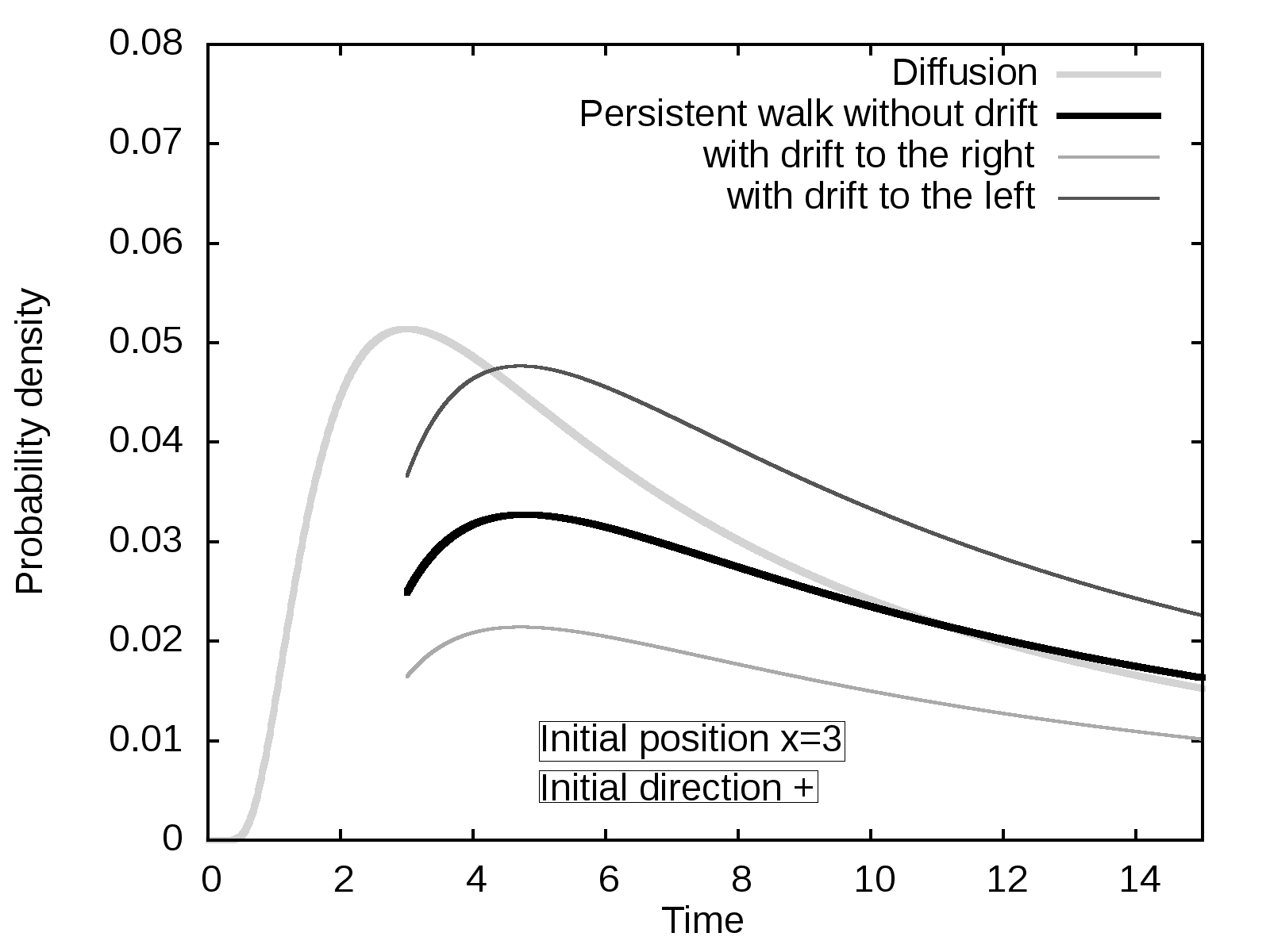}

\includegraphics[width=0.47\textwidth]{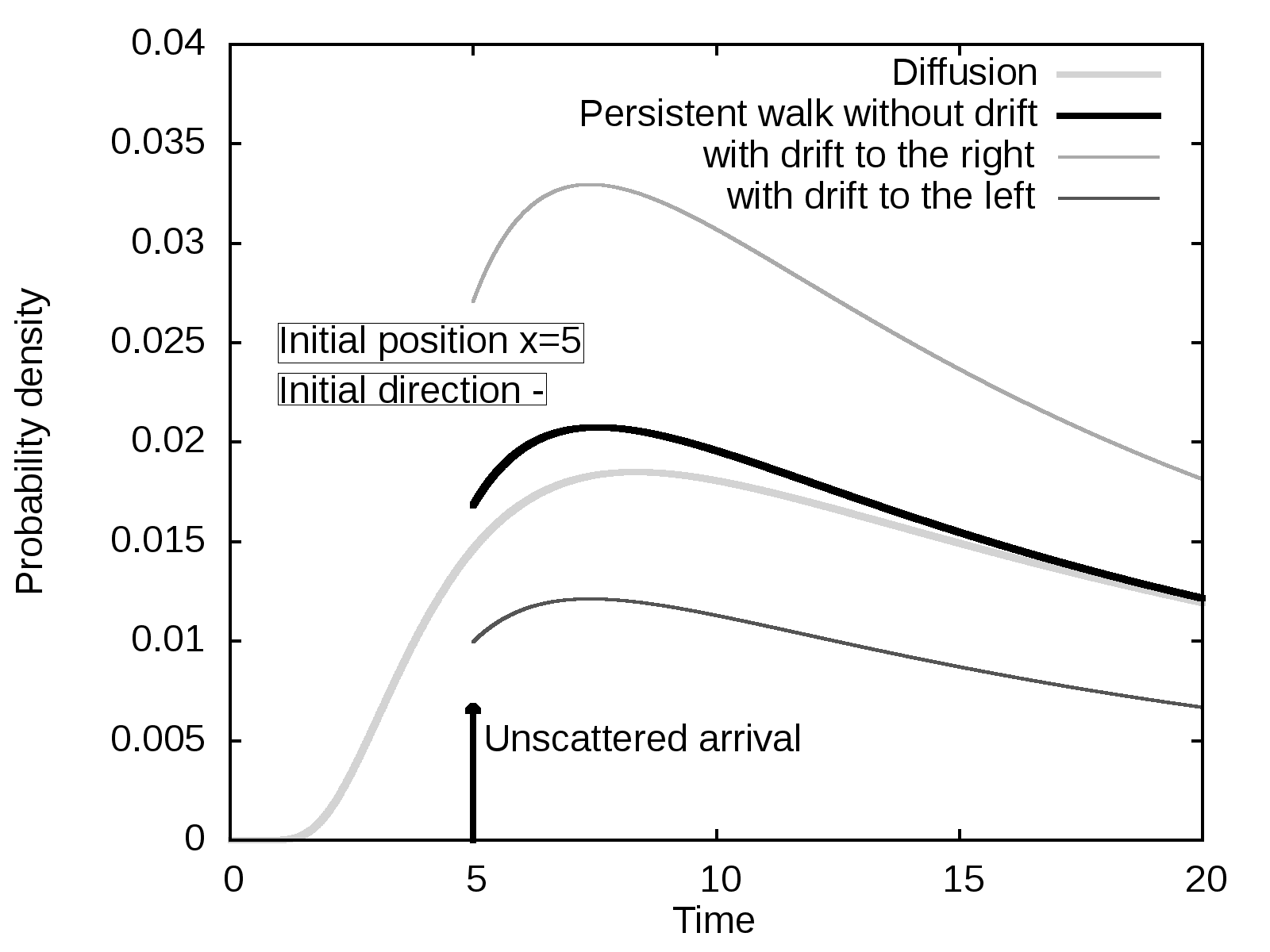}
\includegraphics[width=0.47\textwidth]{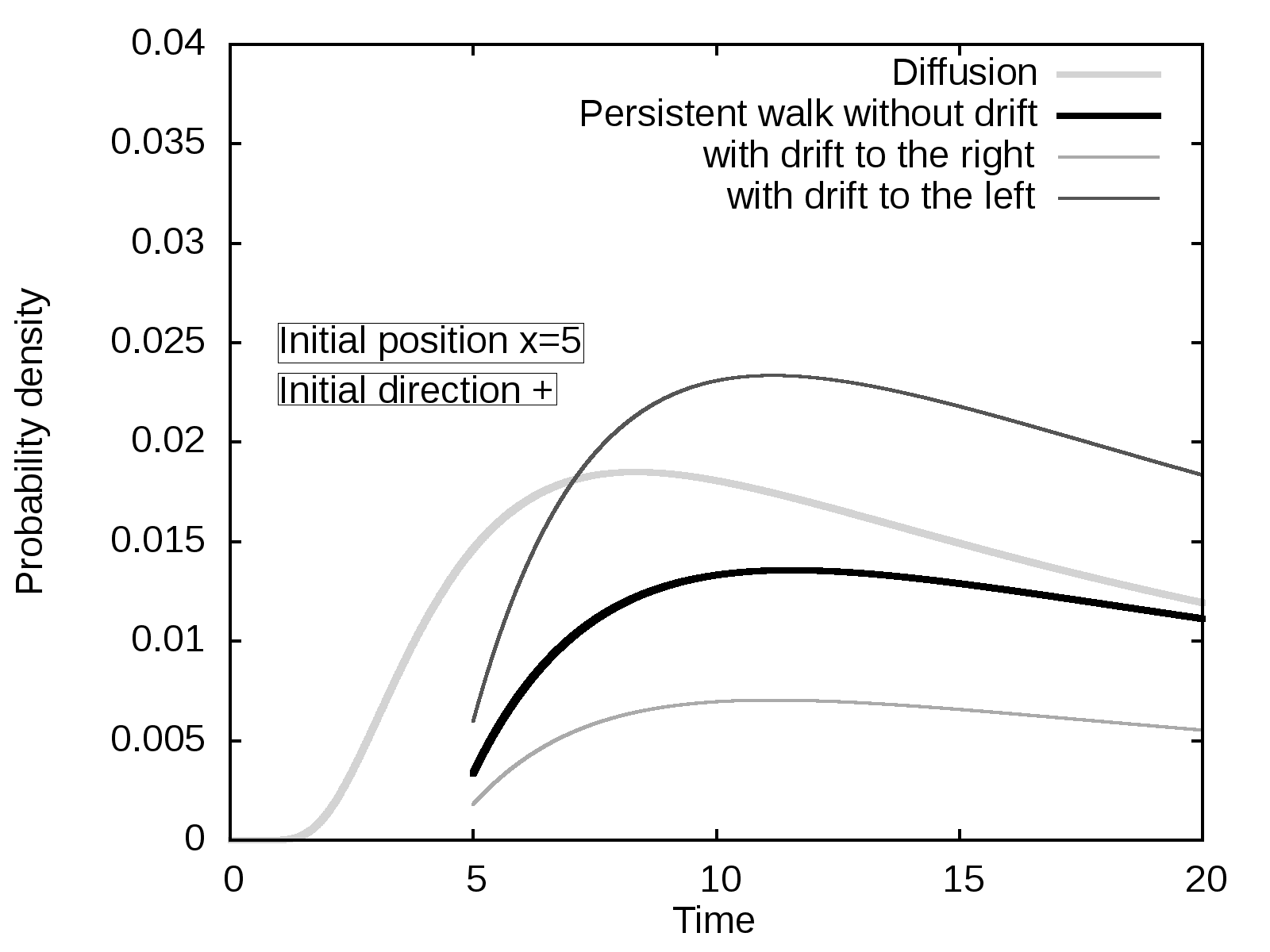}
\caption{\label{figure1d}Comparison between the distribution of
first passage time at the origin for a diffusion process and 
for a persistent random walk. The result shown without drift
is the case $\kappa=0$, while the drifted ones correspond to
the cases $\kappa=0.1$ and $\kappa=-0.1$ for the drift
to the right and the drift to the left respectiveley.
The processes 
start at~$x_0=1$, $3$ and $5$ with initial direction $\mx$ (left) or $\px$ (right).
One observes the ``direct'' unscattered arrival of the persistent 
walk as a peak for initial direction~$\mx$.
The units are fixed by~$c=1$, $\lp=1$ and $D=1=c\lp$ is the diffusion
constant of the diffusion process.}
\end{center}
\end{figure}

The $\px$ case is handled by remarking that the expression~\eqref{siegert}
yields the symmetrical relation
$\TL r(s\,|\,\px)=\TL r(s\,|\,\mx)\big(s+\mu-\sqrt{(s+\mu)^2-1}\big)\lp_-$. 
Invoking the identities \eqref{dxU} and \eqref{u1x} in the Appendix, we get
\begin{equation}
r_z(t\,|\,x_0,\,\px)=\frac2{\lp_+}\frac{x_0}{t+x_0}\gscat[z](-x_0,\,t)
       +\frac{2}{z\lp_+}\frac{t-x_0}{t+x_0}\gev[z](-x_0,\,t).
\label{r+}
\end{equation}
Some examples of first-passage time distributions are displayed in 
Figure~\ref{figure1d}

In the symmetric case $\kappa=0$, the first passage time distribution 
at a trap, or one of two traps, is conveniently provided by the method 
of images. This method can only yield solutions that are 
linear combinations of term of the form $p(x_i,\,t\,|\,\pmx)$ 
where several value of $x_i$ may be used. We deduce from this
remark, that the solutions given by the method of images are solution
of the differential equation (for instance the usual telegrapher's 
equation). As soon as $\kappa\neq0$, 
the results \eqref{r-} and \eqref{r+} are \emph{not} 
solutions of the ATE, which means that the  method of images can't 
yield the previous results. 

\subsection{Phenomenology}
In the case $\kappa\leq0$, the walker will reach the origin almost surely,
but it is interesting to notice that when $\kappa>0$, the walker has a finite 
probability of never reaching the origin, because of the effective drift
due to the mean free path asymmetry. 
The probability that the walker reaches the
origin is called the \emph{origin visit probability} $\porig$. 
The computation of this probability is detailed in
the Appendix. We find, for $\kappa>0$, 
\begin{align}
\porig(x_0,\,\mx)&=
   \int_0^\infty r(t\,|\,x_0,\,\mx)\dd t=
   \e^{-2\kappa x_0},\\
\porig(x_0,\,\px)&=
   \int_0^\infty r(t\,|\,x_0,\,\px)\dd t=
   \frac{\lp_-}{\lp_+}\e^{-2\kappa x_0}.
\end{align}
Let us consider the escape probability $\pesc=1-\porig$ when 
$\kappa$ is close to zero. For $\kappa\leq0$, $\pesc=0$. 
For small $\kappa>0$, the slope of $\pesc$ as a function of $\kappa$
is $2x_0$ for the $\mx$ initial condition and $2(x_0+1)$
for the $\px$ initial condition. This means that 
for small $\kappa>0$, the difference
between the two initial conditions is simply a ``shift length'' equal
to $1=\lp$, and most remarkably independent of $x_0$ : A random
walk starting in the $\px$ direction at $x_0$ has the same
probability of escape as a random walk starting in the direction
$\mx$ at $x_0+\lp$. The parameter~$\kappa$ plays a role similar to
an order parameter in a phase transition.

The average time after which the origin is reached, 
$\torig(x_0\,|\,\pmx)$ is infinite 
in the case $\kappa\geq0$, but it becomes finite in the case $\kappa<0$,
also experiencing a ``transition''. Its value is easily obtained using
the relation
\begin{equation*}
\torig(x_0\,|\,\pmx)=
   \int_0^\infty t\,r_1(t\,|\,x_0,\,\pmx)\dd t= 
  \left(-\vphantom{\lim_x}\derp{}s \TL r_1(s\,|\,x_0,\,\pmx)\right)_{s=0}.
\end{equation*}
We obtain $\torig(x_0\,|\,\px)=
           \torig(x_0\,|\,\mx)+\frac{1}{|\kappa|}=\frac{x_0\mu+1}{|\kappa|}$. 
In this expression we observe an effective fixed delay 
equal to $1/|\kappa|$ for the walker starting
in direction~$\px$ to find itself in the same condition as if it
started in the direction~$\mx$. We may call it a ``flip delay''. 
The average time for a start in the $\mx$ direction is 
proportionnal to $x_0$, showing that the effective movement is
uniform at an effective speed $|\kappa|/\mu$.
In the case $\kappa$, all the higher moments of the first-time passage
are also finite and computed using the same 
technique as for~$\torig$. We can for instance 
compute the fluctuations $\torig[(2)]$ of the
first-time passage at the origin and we find
\begin{equation*}
\torig[(2)](x_0\,|\,\px)=
\torig[(2)](x_0\,|\,\mx)+\frac{\mu}{|\kappa|^3}=
\frac{x_0+\mu}{|\kappa|^3}.
\end{equation*}
We read in this result the fluctuations of the flip delay 
as $\mu/|\kappa|^3$ together with the fluctuations of the uniform 
movement duration $x_0/|\kappa|^3$.

\subsection{A remark concerning the distributions of extrema}
From the statistics of first passage times, it is possible to deduce
the joint probability of the minimum and maximum values of the position~$x$
during a persistent random walk of duration~$t$, as it was first shown
by Masoliver and Weiss~\cite{masoliver1993maximum}.
We start by remarking that the probability that the minimum absciss
reached by the walker during a time $t$ is greater than zero is equal
to the probability that the walker does not reach the origin during 
this time interval :
\[ \proba\left[\text{the minimal absciss reached during}\;[0,\,t] 
    \;\text{is}\;\geq 0 
  \, \vrule\; x_0,\,\pmx\right]
  =1-\int_0^t r_1(t'\,|\,x_0,\,\pmx)\,\dd t'.\]
This setup is equivalent to the situation where the walker starts
from the origin and reaches a negative minimum $-x$ (with $x>0$). 
By derivation with respect to $x$ we find that the spatial 
distribution of the minimum absciss reached by a walker starting
at the origin is
\begin{equation}
m(-x\,|\,t,\,\pmx)=\int_0^t \derp{}x r_1(t'\,|\,x,\,\pmx)\,\dd t'
\qquad (x>0).
\label{m}
\end{equation}
For $x>0$, we have of course $m(x\,|\,t)=0$. 
The distribution of the maximum position is obtained from the same
formula, by remarking that it is the symmetric distribution obtained
with the opposite sign of $\kappa$. Note that for $\kappa\neq0$, 
the expression \eqref{m} cannot be expressed in terms of the
elementary functions $\gscat$ and $\gev$, a numerical approach
is thus required.

\section{First-passage statistics in a finite interval}
The first-passage time distributions we have obtained in the
preceding section have been obtained thank to Siegert's relation,
a technique only valid for a single boundary problem. The 
asymmetric persistent random walk in a finite interval presents
mathematical difficulties with regard to the full first-passage
distribution. It is nonetheless possible to derive some relevant
time integrated 
statistical properties such as the splitting probabilities and the
mean exit time. These quantities can be obtained by very general
methods which realisations do not present any new difficulty here.

We consider the random walk in the interval $[0,\,L]$. 
In the steady state~$p^+$ and $p^-$ are by definition independent of
the time and also independent of the initial condition, position and
direction of propagation. 
The equations \eqref{eq pj} reduce to 
\begin{equation}
\derp{\crt}{x}=0 \qquad \derp{p}{x}=-2\mu j+2\kappa p.
\label{eq pj stationnaire}
\end{equation}
The stationnary current is obviously uniformly equal to $0$ and
the stationnary distribution is immediately obtained, after
normalization, as
\begin{equation}
  p_{\rm ss}(x)=\frac{2\kappa}{\e^{2\kappa L}-1} \,\e^{2\kappa x}.
\end{equation}

\subsection{Splitting probabilities}
We consider the splitting probabilities $q(x\,|\,\pmx)$ that
the walker starting at position $x$ in the direction $\pmx$ reaches
the origin at $x=0$ before ever visiting the other end of the interval at $x=L$. 
In a model where the walker escapes irreversibly from the interval whenever
it reaches any of the two ends, $q$ is the probability that the 
walker escapes from $x=0$. 
Consider a trajectory starting at $x$ in the direction $\px$. It has a 
probability $q(x\,|\,\px)$
of reaching the origin first. After a short time~$\Delta t$, the walker can 
either be still moving in the $\px$ direction or moving in the 
opposite direction and having reached the position $x+\Delta x'$
where $|\Delta x'|\leq c\Delta t$. The probability $q(x\,|\,\px)$ is
therefore equal to
\begin{equation}
q(x\,|\,\px)=\left(1-\frac{c\Delta t}{\lp_+}\right)q(x+c\Delta t\,|\,\px)
              +\frac{c\Delta t}{\lp_-}q(x+\Delta x'\,|\,\mx).
\label{equation separation}
\end{equation}
Taking the first order in $\Delta t$ and proceeding in the same way with the
initial condition $\mx$ 
we obtain the backward Kolmogorov equations
\begin{equation}
\derp{}{x}q(x\,|\,\px)=\derp{}{x}q(x\,|\,\mx)
  =\frac{q(x\,|\,\px)}{\lp_+}-\frac{q(x\,|\,\mx)}{\lp_-}.
\label{back kolmo}
\end{equation}
The boundary conditions for these functions are 
$q(0\,|\,\mx)=1$ and $q(L\,|\,\px)=0$. The solutions of the
equations \eqref{back kolmo} are of the form 
$q(x\,|\pmx)=A_\pm\e^{-2\kappa x}+B_\pm$. The boundary conditions
provide two relations between the integration constants, $A_+\e^{-2\kappa L}+B_+=0$
and $A_-+B_-=1$, while the Kolmogorov equations provide two further
relations $A_+=A_-$ and $B_+/\lp_+=B_-/\lp_-$. This is enough to get the
splitting probabilities
\begin{equation}
q(x\,|\,\pmx)=\frac{\lp_+\e^{-2\kappa x} -\lp_\pm\e^{-2\kappa L}}
           {\lp_+-\lp_-\e^{-2\kappa L}}.
\end{equation}
Interestingly, the difference between these probabilities is uniform
$q(x\,|\,\mx)-q(x\,|\,\px)=2\kappa\e^{-2\kappa L}/(\lp_+-\lp_-\exp(-2\kappa L))$
that we interpret as the probabilistic advantage to start moving
towards the origin.

\subsection{Mean exit time}
Since the walker exits the interval $[0,\,L]$ with probability~1, it is
interesting to know how long it takes on average to exit the interval.
We denote by~$\texit(x\,|\,\pmx)$ the average time taken by a walker
starting from the position~$x$ and moving in the direction~$\pmx$. 
Let us proceed as for the splitting probabilities and consider
a walker located at $x$ moving in the direction $\px$. After 
a time interval $\Delta t$ the mean exit time has been reduced
by $\Delta t$ and the walker can
either be still moving in the $\px$ direction or moving in the
opposite direction and having reached another position $x+\Delta x'$
where $|\Delta x'|\leq c\Delta t$. The mean exit times are related by
\begin{equation}
\texit(x\,|\,\px)=\Delta t+
   \left(1-\frac{c\Delta t}{\lp_+}\right)\texit(x+c\Delta t\,|\,\px)
              +\frac{c\Delta t}{\lp_-}\texit(x+\Delta x'\,|\,\mx).
\label{equation sortie}
\end{equation}
Applying the same reasoning with the initial condition~$\mx$, we 
obtain the backward Fokker-Planck equations
\begin{equation}
\derp{}{x}\texit(x\,|\,\pmx)=
  \frac{\texit(x\,|\,\px)}{\lp_+}-\frac{\texit(x\,|\,\mx)}{\lp_-}\mp1
\label{back fp}
\end{equation}
with the boundary conditions $\texit(0\,|\,\mx)=\texit(L\,|\,\px)=0$. 
The solution of these coupled differential equations follows teh same steps
as the computation of~$q(x\,|\,\pmx)$ and we obtain the expressions
\begin{equation}
\texit(x\,|\,\px)=\frac{(\lp_+)^2(L-x)+(x+\mu)\e^{-2\kappa L}
                                      -(L+\mu)\e^{-2\kappa x}}
                       {\kappa\lp_+-\kappa\lp_-\e^{-2\kappa L}}.
\end{equation}

\section{Concluding remarks}
Recensing all possible symmetry breaking for the one-dimensional 
persistent random walks, we have focused our efforts on the only case leading
new physical results, the case of mean free path anisotropy. This
case is physically relevant in, for instance, microtubule growth.
Statistical quantities such as the Green's function, the generating
function of the number of scattering events, the first-passage 
statistics are obtained from standard computation techniques.

However, one of the first-passage time distributions obtained in the asymmetric 
case, the Equation~\eqref{r+}, cannot be obtained by the simplest,
standard method used in the symmetric case, even though a closed form 
was found. This shows that the asymmetric case $\kappa\neq0$ describes
a new physical situation, not present in the symmetric case. In this
situation the probability of return at the origin is strictly less
than 1, so we may speak of a recurrent-transient transition (second order)
at $\kappa=0$. 
We also remarked that for $\kappa\neq0$ some statistical distributions
do not even have exact analytical forms.
The case $\kappa=0$ is indeed peculiar, because 
$\pesc=0$ and $\torig$ is infinite, whereas in the case where
$\kappa\neq0$ either $\pesc$ or $\torig$
takes a finite value. 

All along the text, we have distinguished the walk by the initial 
direction of propagation~$\px$ or $\mx$ and we have observed that
the results strongly differ depending on it. As Dunkel and H\"anggi
noted, the persistent random walk is not a Markov process \cite{dunkel2009}.
If indeed one takes the initial direction of propagation as a random
variable with $\proba(\px)+\proba(\mx)=1/2$, then the net initial 
current is $j=0$. At a time $t>0$ the current is generally not equal 
to zero, which means that the propagator from time $t$ to a time $t'>t$
does not have the same initial condition for current as at time $t=0$. 
However, we must observe
that the process's construction implies that the Markovian property 
is restored if one considers the directions of propagation separately.
Chapman-Kolmogorov relations follow as
\[ \sum_{\sigma=\px,\mx}\int 
        p^{\beta}(y-x,\,t_1\,|\,\sigma)\,
        p^{\sigma}(x,\,t_2\,|\,\alpha)\;\dd x
  =p^{\beta}(y,\,t_1+t_2\,|\,\alpha), \]
where $\alpha$, $\beta$ and $\sigma$ stand for signs $\px$ or $\mx$
and $p^{\pm}$ functions are given by the Equations~\eqref{p+1} and~\eqref{p-1}.

We have not investigated the situation where $\kappa\neq0$ in the presence
of a speed asymmetry. The position spatial distribution in this situation
follows straightforwardly from this work, but the first-passage time
distributions do not. Even though the effects of both asymmetries separately
are drifts of the average position of the walker along time, these
asymmetries cannot be reconciled into a single one, because the speed
asymmetry is a relativistic change of referential and the 
asymmetry $\kappa\neq0$
cannot be reduced to such a change. The evidence for this impossiblity 
is provided by
the fact that the first-passage time distributions are not solutions
of the ATE for $\kappa\neq0$, as explained in Section~\ref{fp}, although
from the linearity of the change of reference frame, they are solutions
of the ATE in the case of speed asymmetry.

The results of this work may find applications in the domains where the
telegrapher's equation is relevant, extended to the situations where
a scattering asymmetry is present. 
The phenomenology of the first-passage time, introducing the shift length 
and the flip delay between two walkers starting in opposite directions,
may prove useful to interpret experiments in biophysics, optics.


\paragraph{Aknowledgments}
I would like to thank Josh Myers for our fruitful discussions and
Sidney Redner for his nice suggestions.

\appendix
\setcounter{equation}{0}
\renewcommand\theequation{\thesection.\arabic{equation}}


\section{Counting statistics of the $\px$ and $\mx$ walkers}
\label{stat n+/-}
The counting statistics fot the walkers that in the states
$\px$ or $\mx$ at time~$t$ are obtained from the same algebraic
derivations written in Section~\ref{proba}. 

\begin{align}
\TZ P^\pm(z\,|\,x,\,t\,|\,\pmx) & = \frac{\gev[z](x,\,t)}{\gev[1](x,\,t)} 
\label{gen n+}\\
\moy{n^\pm(x,\,t\,|\,\pmx)}     & = \frac{\gscat[1](x,\,t)}{\gev[1](x,\,t)} 
\label{moy n+}\\
\big(\Delta n^\pm(x,\,t\,|\,\pmx)^2\big) & = 
   (t^2-x^2)+2\frac{\gscat[1](x,\,t)}{\gev[1](x,\,t)}
            -\frac{\gscat[1](x,\,t)^2}{\gev[1](x,\,t)^2} 
\label{fluct n+}\\
\TZ P^\mp(z\,|\,x,\,t\,|\,\pmx) & = \frac{\gscat[z](x,\,t)}{\gscat[1](x,\,t)}
\label{gen n-}\\
\moy{n^\mp(x,\,t\,|\,\pmx)}     & = 
   1+(t^2-x^2)\frac{\gev[1](x,\,t)}{\gscat[1](x,\,t)} 
\label{moy n-} \\
\big(\Delta n^\mp(x,\,t\,|\,\pmx)^2\big) & = 
   (t^2-x^2)\left[1-(t^2-x^2)\frac{\gev[1](x,\,t)^2}{\gscat[1](x,\,t)^2}\right]
\label{fluct n-}
\end{align}

From these expressions, we obtain some statistical bounds, using the positivity
of $(\Delta n^\pm)^2$. From the equations \eqref{moy n+} and \eqref{fluct n+}
we get a higher bound for $\moy{n^\pm(x,\,t\,|\,\pmx)}$ and 
from \eqref{moy n-} and \eqref{fluct n-} a higher bound for 
$\moy{n^\mp(x,\,t\,|\,\pmx)}$. Remarking that \eqref{moy n+} and \eqref{moy n-}
provide also the relation $(\moy{n^\pm}-1)\moy{n^\mp}=t^2-x^2$
we have the following bounds 
\begin{align}
1+\frac{t^2-x^2}{1+\sqrt{t^2-x^2}} & \leq 
  \moy{n^\pm(x,\,t\,|\,\pmx)}
  \leq1+\sqrt{1+t^2-x^2}\\
\frac{t^2-x^2}{\sqrt{1+t^2-x^2}} & \leq 
  \moy{n^\mp(x,\,t\,|\pmx)}\leq
   1+\sqrt{t^2-x^2}.
\end{align}

\section{The functions~$\gscat[z]$ and $\gev[z]$ and their properties}
\setcounter{equation}{0}
\subsection{Definitions}
\begin{align}
\gscat[z](x,\,t)&=z\frac{\e^{\kappa x-\mu t}}{2}
  I_0\left(z\tx\right)\Theta(t-|x|)\\
\gev[z](x,\,t)&=z\frac{\e^{\kappa x-\mu t}}2\frac{I_1\left(z\tx\right)}\tx
 \Theta(t-|x|)
\end{align}

\subsection{Transforms}
\begin{align}
    \frac{z}{(s+\mu)^2+(k-\ii\kappa)^2-z^2}&=
\TFLgscat[z](k,\,s)\\
z\frac{\e^{\kappa x-|x|
    \sqrt{(s+\mu)^2-z^2}}}{\sqrt{(s+\mu)^2-z^2}}&=
2\TLgscat[z](x,\,s)\\
\e^{\kappa x-|x|\sqrt{(s+\mu)^2-z^2}}&=
-2|x|\,\TLgev[z](x,\,s)+\e^{\kappa x-(s+\mu)\,|x|}
\label{U}
\end{align}
\begin{multline}
\e^{\kappa x-|x|\sqrt{(s+\mu)^2-z^2}}\sqrt{(s+\mu)^2-z^2}=\\
 2z\TLgev[z](x,\,s)-2\kappa x\TLgev[z](x,\,s)
+2x\derp{\TLgev[z]}{x}(x,\,s)+\sign(x)(s+\mu)\e^{\kappa x-(s+\mu)|x|}
\label{dxU}
\end{multline}


\subsection{Derivatives}
\begin{align}
\derp{\gscat[z]}t(x,t)&=-\mu\gscat[z](x,t)+ zt\;\gev[z](x,t) 
               +\frac 12\dirac(t-x)\e^{-t/\lp_+}
               +\frac 12\dirac(t+x)\e^{-t/\lp_-},
\label{u0t}\\
\derp{\gscat[z]}x(x,t)&=\kappa\gscat[z](x,t)- zx\;\gev[z](x,t)
               -\frac 12\dirac(t-x)\e^{-t/\lp_+}
               +\frac 12\dirac(t+x)\e^{-t/\lp_-},
\label{u0x}\\
\derp{\gscat[z]}z(x,t)&=\frac{1}{z}\gscat[z](x,t)+(t^2-x^2)\gev[z](x,t),
\label{u0z}\\
\derp{\gev[z]}t(x,t)&=-\mu\gev[z](x,t)
              +\frac{zt}{t^2-x^2}(\gscat[z](x,t)-2\gev[z](x,t))
              +\frac 14\dirac(t-x)\e^{-t/\lp_+}
              +\frac 14\dirac(t+x)\e^{-t/\lp_-},
\label{u1t}\\
\derp{\gev[z]}x(x,t)&=\kappa\gev[z](x,t)-\frac{zx}{t^2-x^2}
                          (\gscat[z](x,t)-2\gev[z](x,t))
              -\frac 14\dirac(t-x)\e^{-t/\lp_+}
              +\frac 14\dirac(t+x)\e^{-t/\lp_-},
\label{u1x}\\
\derp{\gev[z]}z(x,t)&=\gscat[z](x,t).
\label{u1z}
\end{align}
We have used the identity $I_2(x)=I_0(x)-\frac2xI_1(x)$.

\section{Probability of reaching the origin}
\setcounter{equation}{0}
Let us first compute $\porig(x_0,\,\mx)$ by remarking
that it is the value of the Laplace transform of $r$ at $s=0$ :
\[ \begin{split}
\porig(x_0,\,\mx)&=\int_0^\infty r_1(t\,|\,x_0,\,\mx)\dd t
=\int_0^\infty \left[2x_0 u_1(-x_0,\,t)
     +\dirac(t-x_0)\e^{-x_0/\lp_-}\right]\dd t\\
&=2x_0 \,\TL u_1(-x_0,\,0)+\e^{-x_0/\lp_-}=\e^{-\kappa x_0-x_0|\kappa|}-
\e^{-\kappa x_0-\mu x_0}+\e^{-x_0/\lp_-}=\e^{(-\kappa-|\kappa|)x_0}
\end{split}\] 
and observe that the probability is equal to $1$ for $\kappa<0$
and
strictly $<1$ for $\kappa>0$. The computation of
 $\porig(x_0,\,\mx)$ requires more work :
\[ \begin{split}
\porig(x_0,\,\px)&=\int_0^\infty r_1(t\,|\,x_0,\,\px)\dd t\\
&=\int_0^\infty \frac1{\lp_+}\left[\frac{x_0}{t+x_0}\e^{-\kappa x_0-\mu t}
   I_0\left(\sqrt{t^2-x_0^2}\right)
   +\frac{t-x_0}{t+x_0}\e^{-\kappa x_0-\mu t}
   \frac{2I_1\left(\sqrt{t^2-x_0^2}\right)}{2\sqrt{t^2-x_0^2}}\right]\dd t
\end{split}\]
Using the identity $\frac{x_0}{t+x_0}=\frac{1}{2}-\frac12\frac{t-x_0}{t+x_0}$
this expression rewrites
\[ \begin{split}
\porig(x_0,\,\px)&=
 \frac{\e^{-\kappa x_0}}{\lp_+}
  \int_0^\infty \e^{-\mu t} \frac12\frac{t-x_0}{t+x_0}
  \left[-I_0\left(\sqrt{t^2-x_0^2}\right)
  +2\frac{I_1\left(\sqrt{t^2-x_0^2}\right)}{\sqrt{t^2-x_0^2}}\right]\dd t
\;\cdots\\
&\quad\qquad\cdots 
 +\frac{\e^{-\kappa x_0}}{2\lp_+}\int_0^\infty
   \e^{-\mu t}I_0\left(\sqrt{t^2-x_0^2}\right)\dd t\\
 &=\frac{\e^{-\kappa x_0}}{2\lp_+}\frac{\e^{-x_0\sqrt{\mu^2-1}}}{\sqrt{\mu^2-1}}
   -\frac{\e^{-\kappa x_0}}{2\lp_+}\int_0^\infty\e^{-\mu t}\frac{t-x_0}{t+x_0}
     I_2\left(\sqrt{t^2-x_0^2}\right) \dd t.
\end{split}\]
The integral is the Laplace transform 4.17(11) in the table~\cite{erdelyi1}
for the case $\nu=2$ such that we obtain
\[ 
\porig(x_0,\,\px)=
\frac{1}{2\lp_+}\frac{\e^{(-\kappa-|\kappa|)x_0}}{|\kappa|}
 \left(1-\left(\mu-\sqrt{\mu^2-1}\right)^2\right)
 =\frac{1}{2\lp_+}\frac{\e^{(-\kappa-|\kappa|)x_0}}{|\kappa|}
 \left(1-(\mu-|\kappa|)^2\right).
\]
If $\kappa<0$ we have $\porig(x_0,\,\px)=
  \frac{\e^{(-\kappa+\kappa)x_0}}{2(\mu+\kappa)(-\kappa)}
  \left(1-\mu^2-\kappa^2-2\kappa\mu\right)=1$.
In the case $\kappa>0$ we finally obtain
$ \porig(x_0,\,\px)=
  \frac{\e^{(-\kappa+\kappa)x_0}}{2(\mu+\kappa)\kappa}
  \left(1-\mu^2-\kappa^2+2\kappa\mu\right)=
  \e^{-2\kappa x_0}\frac{\mu-\kappa}{\mu+\kappa}<1$.


\bibliography{tr1d}
\bibliographynote{tr1d}

\end{document}